\documentclass[aps,pra,preprint]{revtex4-1}
\usepackage[latin1]{inputenc}
\usepackage[english]{babel}
\usepackage{latexsym}
\usepackage{graphicx}
\usepackage{amsmath}
\usepackage{natbib}
\newcommand{\au}{a$_0$}
\newcommand{\rp}{$\langle r_p\rangle$}
\newcommand{\re}{$\langle r_e\rangle$}
\newcommand{\repCM}{$\langle r_{e-p}^{CM}\rangle$}

\begin{document}

\title{Matter-positronium interaction: An exact diagonalization study of
the He atom - positronium system}
\author{A. Zubiaga}
\email{asier.zubiaga@aalto.fi}
\author{F. Tuomisto}
\author{M. J. Puska}
\affiliation{Department of Applied Physics, Aalto University, P.O. Box 11100, FIN-00076 Aalto Espoo, Finland}

\begin{abstract}
The many-body system comprising a He nucleus, three electrons, and a positron
has been studied using the exact diagonalization technique. The purpose
has been to clarify to which extent the system can be considered as a
distinguishable positronium (Ps) atom interacting with a He atom and,
thereby, to pave the way to a practical atomistic modeling of Ps states
and annihilation in matter. The maximum value of the distance between the
positron and the nucleus is constrained and the Ps atom at different distances
from the nucleus is identified from the electron and positron densities,
as well as from the electron-positron distance and center-of-mass
distributions. The polarization of the Ps atom increases as its distance from 
the nucleus decreases. 
A depletion of the He electron density, particularly large at low density values, has been observed.
The ortho-Ps pick-off annihilation rate calculated as 
the overlap of the positron and the free He electron densities has to be corrected for the 
observed depletion, specially at large pores/voids.
\end{abstract}


\maketitle

\section{Introduction}
The o-Ps atom is the bound state 
of an electron and a positron with total spin $S$ = 1. In vacuum
o-Ps has a relatively long lifetime of 142 ns because its annihilation 
via the fast two-gamma channel is prohibited by the conservation of the
angular momemtum. 
Annihilation through the two-gamma channel with an electron of the matter having an opposite spin is possible 
when o-Ps interacts with matter. 
The resulting pick-off annihilation depends on the overlap of the o-Ps with the electron density of the
matter and it can reduce the positron lifetime remarkably~\cite{BookChapter_Mogensen}. 
In metals and semiconductors the electron density is enhanced around 
the positron  due to the attractive electron-positron interaction. 
This means that a Ps-like quasiparticle is created by the short-range 
screening of the positron by the electron cloud. In 
molecular matter and in some insulators such as SiO$_2$, 
the electron density is low in the interstitial regions and
a Ps atom can be formed~\cite{BookChapter_Jean}. In this case 
the electron accompanying the positron hinders efficiently further
screening of the positron by electrons in the matter.
On the other hand, there is a finite 
overlap between the positron and the electrons of the material
manifesting itself as the repulsive Pauli interaction and as
the pick-off annihilation of o-Ps. 

The pick-off annihilation lifetime spectroscopy of orto-Positronium (o-Ps) has a rather 
unique role as a method capable to study open volumes in soft or porous 
materials in which Ps is formed. 
Size and density distributions 
of nanometre-scale voids in polymers~\cite{JPS_Uedono}, porous SiO$_2$
\cite{PRB_Nagai, APL_Liszkay}, and biostructures~\cite{JPCBL_Sane, JPCBL_Dong} have been 
estimated by measuring the pick-off annihilation rate of o-Ps. However, in order to make a reliable quantitative analysis  
of the experimental results, one should be
able to model o-Ps states and annihilation in these structures with
predictive power. 

In contrast to an atom where 
the nuclear degrees of freedom can, to a good approximation, 
be treated classically, 
both the electron and the positron in the Ps atom
are light quantum-mechanical particles and 
the non-adiabatic correlation effects have to be taken into account.
For positrons in dense materials, such as
metals and semiconductors, a good starting point is a separate calculation
of the quantum-mechanical state of a single positron~\cite{RMP_Puska}. 
In practice, this calculation is based on the results of many-body theories 
for a delocalized positron in a homogeneous electron gas.
The nature of the Ps atom would call for a quantum-mechanical many-body 
calculation of the interacting electron-positron system in the Coulomb 
field of nuclei. In the case of materials systems, this is clearly beyond
the present computer capacity. An attractive starting point
for modeling Ps is to consider it as a distinguishable particle 
also when it is interacting closely with matter. This is 
in line with the view of a long-living o-Ps
interacting with matter and annihilating through a pick-off process.

The above-mentioned model of the Ps atom interacting as a single particle 
with matter finds support also from experimental and theoretical 
studies of Ps scattering off atoms and light molecules.
Recently, the scattering of Ps off noble gas atoms and light 
molecules has been measured using a slow mono-energetic Ps 
beam~\cite{Science_Brawley}. The
scattering properties of Ps have been seen to be dominated by the
repulsive electron-electron Pauli interaction at short distances. 
At long distances the van der Waals
attraction may play a role, especially in interactions with molecules
of high polarizability. 

The positronium Hydride (HPs) bound state has been observed experimentally 
by Schader et al.~\cite{PRL_Schrader} and its 
binding energy has been estimated to be around 1.1~eV (0.04~Ha). 
The bound states of Ps with several small atoms have been studied theoretically using
Exact Diagonalization techniques. 
The most carefully studied system is HPs~\cite{JP_Ryzhikh,PRA_Mitroy3,PRA_Bubin}, which has been found to form 
a bound state with a binding energy of 0.039 Ha, in good agreement
with the experimental value. LiPs~\cite{JP_Ryzhikh,JAMS_Mitroy,JPB_Mitroy1} and NaPs~\cite{JP_Ryzhikh,JPB_Mitroy1} systems
have also been predicted to form bound states with binding energies
of 0.0123 Ha and 0.0084 Ha, respectively. The calculations show that 
at large distances the positron overlaps with an extra electron forming 
a free Ps-like cluster, while, when the positron density overlaps with 
that of the electrons inside the atom the Ps-like character is lost.
In the case of closed-shell noble gas atoms the strong 
electron-electron Pauli repulsion dominates the atom-Ps interaction 
and the formation of bound states is prevented. 
Thus far, only a small 
number of different atoms have been considered in this kind of accurate 
calculations. However, it is already clear that the atoms able to trap Ps have 
high electron affinities and open-shell electronic structures~\cite{JPB_Mitroy}. 

Mitroy and Ivanov have studied the scattering 
of o-Ps off rare-gas atoms and alkali metal ions.~\cite{PRA_Mitroy2,PRL_Mitroy}
They determined the scattering lengths for wave vectors up to 
0.5~$a_0^{-1}$ corresponding to the Ps energy of $\sim$1.7 eV.   
Besides rare-gas atoms, also molecules have small electron affinities. 
They typically have HOMO-LUMO gaps of up to 1 Ha
and their electrons are well localized in molecular orbitals. 
In polymers and biomolecular materials, the main cohesive interaction between the
molecules is the van der Waals interaction and their molecular orbitals 
are not greatly affected. Individual molecules still show HOMO-LUMO gaps 
and the repulsive Pauli interaction between electrons in neighboring 
molecules plays a role. On the other hand, the van der Waals interaction 
between the molecules and Ps can also be strong, especially for molecules 
of high polarizability. 

A model widely used to study Ps in small pores and voids ($<$ 2 nm)
is that developed by Tao and Eldrup~\cite{JCP_Tao,CP_Eldrup}. The semi-empirical model 
relates the Ps lifetime and the pore/void radius. 
Ps is considered as a single particle and the pore/void is modeled using 
a spherical, cubic or elliptic infinite potential well. 
The electron density of the material enters the void and 
forms a layer of the thickness $\Delta$R on its surface. The pick-off 
annihilation rate is calculated from the overlap of the positron density
with the electron density of the material. The parameter of the model, 
$\Delta$R, is not well known for all the materials and its chemical 
dependence cannot be described. In addition, the model assumes compact pores 
inside the material. The model has been extended to larger pores including the
annihilation with excited particle-in-a-box states distributed according 
the Boltzmann distribution at a finite temperature.~\cite{JPCB_Dull}

Schmitz and M\"uller-Plathe~\cite{JCP_Schmitz} introduced an atomistic model to
describe Ps states and annihilation at voids in polymers. The potential
landscape for Ps is based on the superposition of interactions between
the Ps atom and individual atoms. These interactions, in turn, include
the long-range van der Waals interaction based on the polarizabilities
of the Ps atom and the atoms in the polymer and the short-range Pauli repulsion
between the electron in the Ps atom and the electrons of the material. 
The latter interaction is described by fitting to experimental scattering
cross sections. The Ps wave function corresponding to a given finite 
temperature is solved for by the path integral Monte-Carlo method and 
thereafter the pick-off annihilation rate is calculated as the overlap of 
the positron density with the electron density which is also obtained as 
a superposition over the atoms.

Our final goal is the atomistic modeling of a Ps atom interacting with 
molecular materials so that the necessary approximations made and
the model parameters chosen are based on ab-initio results.
Studying theoretically the HePs system will help 
to understand several aspects in the properties of Ps also in 
molecular solids and liquids. This is because in molecules electrons 
are also in a closed shell configuration and a HOMO-LUMO gap sets 
the minimum energy necessary for an extra electron (of the Ps atom) 
to enter into the electron cloud of the molecule.

The similarities between molecules with
HOMO-LUMO gaps and closed-shell rare gas atoms in mind, 
we have calculated the many-body wavefunction and
the total energy of the unbound HePs system by 
the exact diagonalization technique using a explicitly correlated 
Gaussians basis optimized by a stochastic variational method (ECG-SVM). 
In order to study the interaction of a 
Ps and a He atom at finite distances, the mean distance of 
the positron from the nucleus has been constrained. As a result, 
a set of configurations has been considered with the nucleus-positron mean
distance ranging between 1.87~\au{} and 91.90~\au{}. Electron
and positron densities, interaction energies and annihilation rates
have been calculated and analysed in order to clarify to which
extent the picture of the distinguisable Ps particle can be applied.

\section{Stochastic variational method and the modeling of the \MakeUppercase{H}\MakeLowercase{e}\MakeUppercase{P}\MakeLowercase{s} system}
We have described the $N$-particle system comprising the heavy nucleus 
(treated as a single particle), 
light electrons and the positron by the non-relativistic Hamiltonian
\begin{eqnarray}
\widehat{H} = \sum_i\frac{\vec{p}^2_i}{2m_i} - T_{cm} +
\sum_{i<j}\frac{q_iq_j}{4\pi\epsilon_0r_{ij}},
\end{eqnarray}
where $\vec{p}_i$, $m_i$, and $q_i$, are the momenta, masses, and charges 
of the particles, respectively, $r_{ij}$ is the distance between the 
$i^{\rm th}$ and $j^{\rm th}$ particles, and $T_{cm}$ is the center-of-mass (CM) 
kinetic energy. The wavefunction is written as a linear combination of 
properly antisymmetrized explicitly correlated Gaussian 
(ECG)~\cite{PRC_Varga} functions as 
\begin{widetext}
\begin{eqnarray}
\Psi = \sum_{i=1}^s c_i\ \psi_{SMs}^i(\vec{x},A^i) = \sum_{i=1}^s
c_i\ {\displaystyle \mathcal{A} } \left \{ 
\begin{array}{c}
\exp\left( 
-\frac{1}{2}\sum_{\mu,\nu=1}^{N-1}A^i_{\mu\nu}\vec{x}_{\mu}\vec{x}_{\nu}
\right) 
\otimes\chi_{SMs}
\end{array} 
\right \},
\end{eqnarray}
\end{widetext}
where $\mathcal{A}$ is an antisymmetrization operator and $\chi_{SMs}$ 
is the spin eigenfunction with $\hat{S}^2\chi_{SMs}=S(S+1)\hbar^2\chi_{SMs}$ 
and $\hat{S}_z\chi_{SMs}=M_S\hbar\chi_{SMs}$. We have considered only the 
spherically symmetric total angular momentum $L=0$ states. The mixing 
coefficients $c_i$ are obtained by diagonalizing the Hamiltonian matrix 
$<\psi_{SMs}^i\vert \widehat{H} \vert \psi_{SMs}^j>$. The nonlinear
coefficients $A^i_{\mu\nu}$ are randomly generated and a new $A^i_{\mu\nu}$
is kept only if the update lowers the total energy of the system.
Using the Jacobi coordinate set \{$\vec{x}_1$,...,$\vec{x}_{N-1}$\}, 
with the reduced mass of the $i^{\rm th}$ coordinate as 
$\mu_i$ = $m_{i+1}\frac{\sum_{i=1}^im_i}{\sum_{i=1}^{i+1} m_i}$, allows 
for a straightforward separation of the CM movement and therefore 
the total number of ``particles'' (degrees of freedom) treated is $N-1$. 

The stochastic variational method (SVM) is suitable for calculating
many-body wavefunctions, such as $\Psi$ in Eq. (2), for small 
systems.~\cite{PRC_Varga} 
The number of particles is limited by the present algorithms and 
computer capacity typically to less than seven particles. 
The ECG basis comprises in our calculations 600 functions. The matrix 
elements between the basis functions are computed analytically 
in an efficient manner. In order to keep the size of the ECG basis small, 
the basis set is optimized by choosing the most appropriate non-linear
parameters $A^i_{\mu\nu}$ for the Gaussian functions using a stochastic 
search in the parameter space. The large number of parameters to optimize, 
up to tens of thousands, prevents the use of direct search methods. 
The best non-linear parameter for each pair is chosen after 25 trials within each optimization cycle 
and 20 optimization cycles are done for each basis function. 
The trial values are obtained choosing values between a
certain minimum and maximum using a random number generator. 

SVM can be also used to study unbound systems setting boundaries to the possible values 
of the non-linear parameters $A^i_{\mu\nu}$, i.~e. increasing the minimum value 
the particles can be prevented to separate to the infinity~\cite{PRA_Mitroy2}.
The configurations with the lowest energy are searched for within the available 
parameter space, i.e., the energy is minimized within a constrained parameter space. 
In order to model the interaction between He and the Ps atom at different distances
it is enough to set a constraint only to the non-linear parameter 
$A^i_{\mu\nu}$ corresponding to the nucleus-positron pair. Thus, the size of 
the parameter space available for optimizing the wavefunction is not seriously
restricted. 
The nucleus-positron distance parameter $l_{He-p}$ corresponding 
to the  maximum of the non-linear parameter, $A^i_{He-p}=(1/l^i_{He-p})^2$, 
is varied between 1 and 50~\au{}. 
The wavefunction and the total energy of the HePs system are then 
calculated for each constrained He-Ps distance. 
Due to practical reasons, the nucleus-electron distance parameter 
$l_{He-e}$ also has to be limited. We have checked that using $l^i_{He-e}$~=~200~\au{}
gives structures and energies well-enough converged for our discussion below. 

The kinetic energy ($\langle T\rangle$) and the potential energy
($\langle V\rangle$) of a system interacting through Coulomb
potentials are related by the Virial theorem 2$\langle
T\rangle$=-$\langle V\rangle$. Any deviation from this relation is due
to an inadequate basis function set. Optimizing the basis set the
virial coefficient $(\frac{2\langle T\rangle}{\langle V\rangle}+1)$
decreases in magnitude. 

In order to study the many-body wavefunction of the system, the positron 
($\rho_{p}(r) = \langle\Psi |\delta(\vec{r}_p-\vec{r}_N-\vec{r})|\Psi\rangle$) 
and electron 
($\rho_{e}(r) = \sum_{i=1}^{N_e}\langle\Psi |\delta(\vec{r}_{i,e}-\vec{r}_N-\vec{r})|\Psi\rangle$) 
density distributions are plotted. 
Above, $\vec{r}_N$ is the position vector of the nucleus and the index $i$ runs over all the electrons $N_e$. 
The e-p pair correlation function is calculated as
\begin{equation}\label{eq_ep}
\rho_{e-p}(\vec{r}) = \sum_{i=1}^{N_e}\langle\Psi |\delta(\vec{r}_i-\vec{r}_p-\vec{r})|\Psi\rangle
\end{equation}

The distribution of the electron-positron (e-p) CM is defined as
\begin{equation}
\rho_{ep}^{CM}(\vec{r}) = \sum_{i=1}^{N_e}\langle\Psi |\delta(\vec{r}_i^{CM}-\vec{r})|\Psi\rangle
\end{equation}
where $\vec{r}_i^{CM}=(\vec{r}_{i,e}+\vec{r}_p)/2$ is the center 
of mass of the positron and the $i^{\rm th}$ electron and $i$ runs 
over all the $N_e$ electrons in the HePs system. 
The calculated distributions do not have any angular dependency and 
the radial distributions are the solid-angle averaged distributions multiplied 
by the radius squared. 

The two electrons of the He atom are prepared in the $S=0$ state while the electron and 
the positron of the o-Ps are in the $S=1$ state. 
The total spin angular momentum of the system is, then, $S=1$. 
For the pick-off annihilation rate $\Gamma^{po}$, the $S=0$ component of the $i^{\rm th}$electron-positron 
pair is selected through the $\widehat{P}^{i-p}_{S=0}$ spin projector, i.e., 
\begin{equation}\label{gamma_eq}
\Gamma^{po} = 4\pi r_0^2 c \sum_{i=1}^{N_e} \langle \Psi|\delta(\vec{r}_i-\vec{r}_p)\widehat{P}^{i-p}_{S=0}|\Psi\rangle 
\end{equation}
where $r_0$ is the classical electron radius and $c$ is the speed of light. 
The summation runs over all the $N_e$ electrons. 

\section{Results \& Discussion}
\begin{figure}[ht]
\begin{center}
\includegraphics[width=8.5cm]{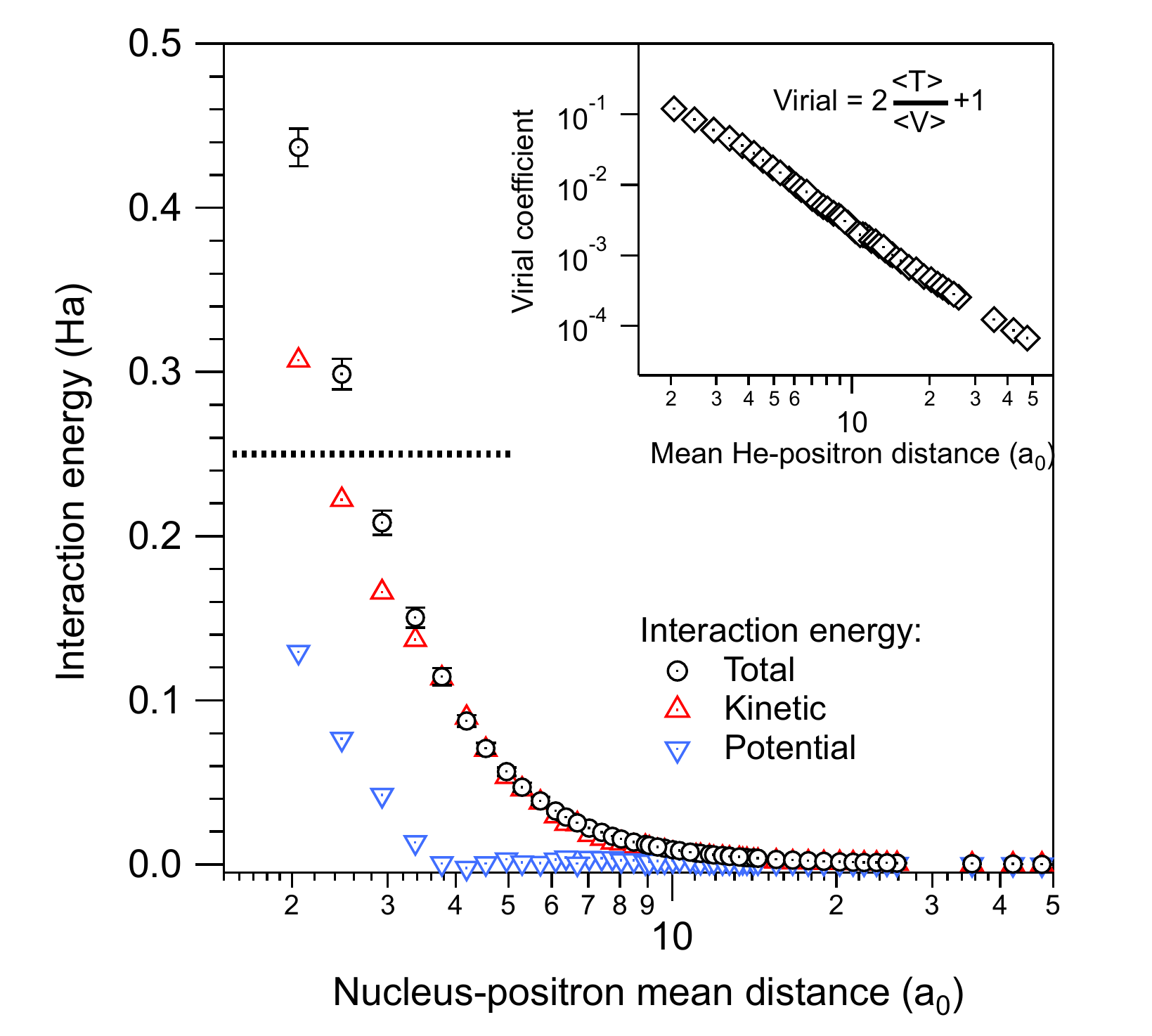}
\caption{Total interaction energy and its kinetic and potential energy 
contributions as a function of \rp{}. 
The horizontal dotted line denotes the energy above which the Ps atom 
is unstable against dissociation.
The inset shows the log-log plot of the virial coefficient versus \rp{}.
}
\label{fig6}
\end{center}
\end{figure}

Before discussing the main topic of this work, the identification of 
the Ps atom within the HePs system, we will shortly consider the 
energetics of the HePs system.
The interaction energy $E_I$ of HePs is defined as the difference between 
the total energy of the interacting system and the sum of the energies 
$E_{He}$ and $E_{Ps}$ of the isolated He and Ps atoms, respectively, i.e., 
$E_I=E_{HePs}-E_{He}-E_{Ps}$. 
$E_{Ps}$=-0.24999999999~Ha and $E_{He}$ = -2.903693749~Ha have 
been obtained from SVM calculations with a basis of 56 functions (virial coefficient = 
3.0$\times$10$^{-11}$) and 600 functions (virial coefficient = 1.8$\times$10$^{-10}$), 
respectively.
The interaction energy is shown in figure~\ref{fig6} as a function of the 
nucleus-positron mean distance \rp{}. $E_I$ is positive 
and approaches zero for large \rp{} values. 
For \rp{} smaller than 2.5~\au{}, $E_I$ is larger than 
$E_{Ps}$. 
Our calculations do not show the formation of the bound 
He-e$^+$ system. He can bind a positron only when it is in the 
$S=3$ state~\cite{PRA_Mitroy} and its energy is 0.4 Ha above the 
the highest energy we have considered for the HePs system. 
It should be noticed that, already at 5~\au{}, the interaction energy is clearly 
above the energy of a thermalized Ps. 

The error of the interaction energy $\Delta$E has been estimated as 
the total energy decrease after an optimization loop of the function basis. 
For the optimization loop the best non-linear parameter has been chosen 
for each pair after 25 trials. 
This procedure has been repeated 20 times for each basis function.
For the most confined systems, \rp{}= 2~\au{}, $\Delta$E = 0.01~Ha and 
the virial coefficient is $\sim$ 0.1. 
The confinement method shrinks the parameter space available to optimize the basis. 
In addition, the many-body wavefunction has to include the strong correlations
happening in confined systems. 
Accordingly, the convergence is slow and the virial coefficient is large. 
For \rp{} = 48~\au{}, the error and the virial coefficient decrease to 
$\Delta$E = 2$\times$10$^{-5}$~Ha and 7$\times$10$^{-5}$, respectively. 

Figure~\ref{fig6} also shows the decomposition of the interaction energy 
to the kinetic and potential energy contributions. They are defined similarly
to the total interaction energy.
E$_{Kin}$ is the main contribution to the interaction energy 
at distances shorter than 10~\au{}, mostly due to the confinement of Ps and its small mass. 
E$_{Pot}$ is always positive and 
at very short distances ($<$ 4~\au{}) it grows very fast. 

\begin{figure}[ht]
\begin{center}
\includegraphics[width=8.5cm]{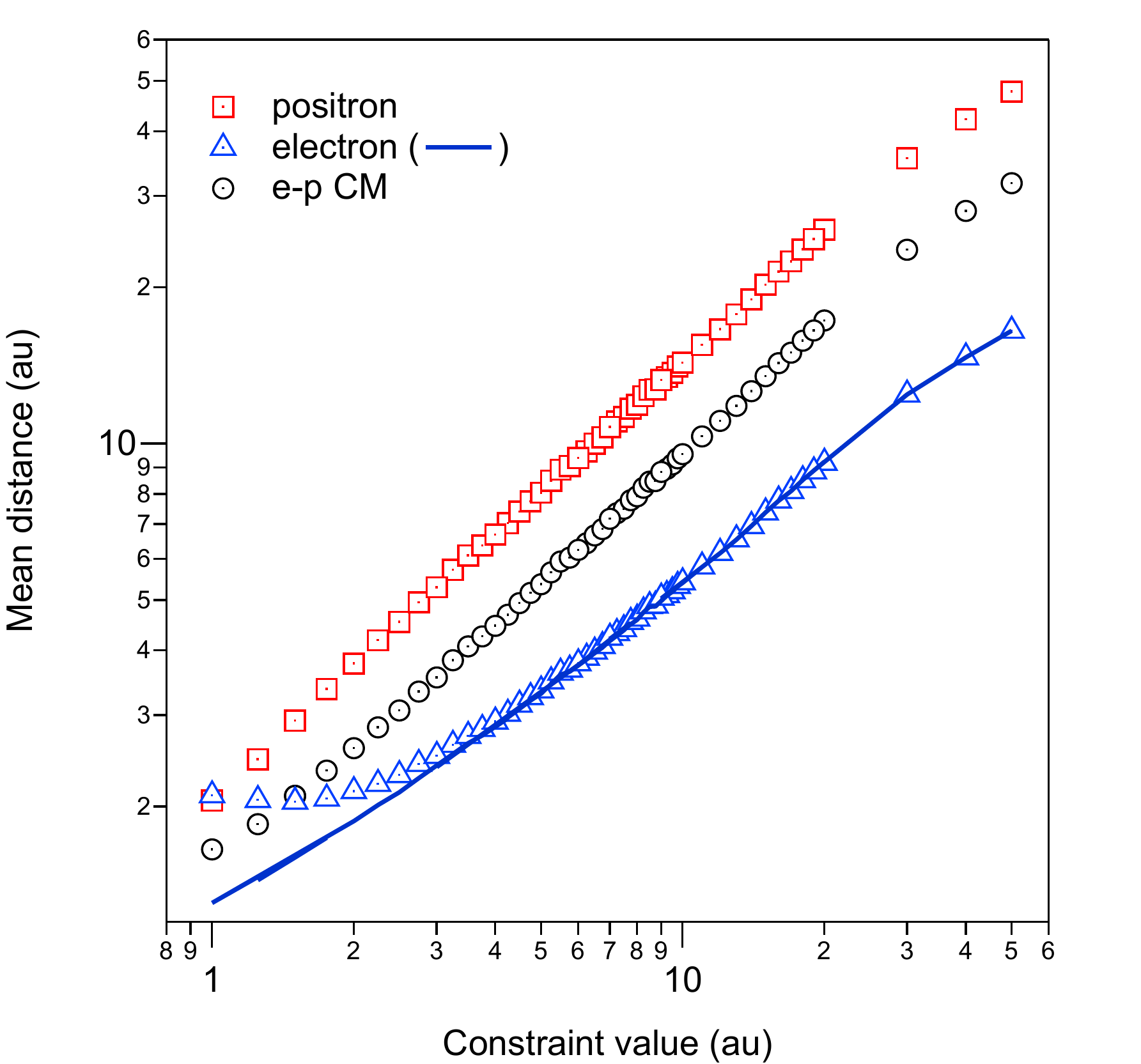}
\caption{(Color online) Positron (red squares), electron (blue triangles) 
and electron-positron CM (black circles) mean distances from the He 
nucleus for different calculated configurations versus the nucleus-positron 
distance parameter $l_{He-p}$. The electron mean distance (blue line) obtained from 
the unpolarized Ps model is also given.} 
\label{fig5}
\end{center}
\end{figure}

The calculated mean distances of electrons \re{}, the positron \rp{} and the 
electron-positron CM \repCM{} from the He nucleus are shown in Fig.~\ref{fig5}
as a function of the nucleus-positron distance parameter $l_{He-p}$ 
corresponding to the confinement of the system.
The mean distances grow monotonously because HePs is ultimately unbound. 
The fact that \re{} increases with \rp{} is an indication of the formation of Ps. 
At $l_{He-p}$ $<$ 5~\au{}, \re{} saturates to a value of 2~\au{}
because the electron of Ps does not penetrate into the He core. 
On the average, it stays at $\sim$ 4~\au{} from the nucleus (see figure~\ref{fig1}). 
On the other hand, \repCM{} continues to decrease because the positron keeps being confined. 

\begin{figure}[ht]
\begin{center}
\includegraphics[width=8.5cm]{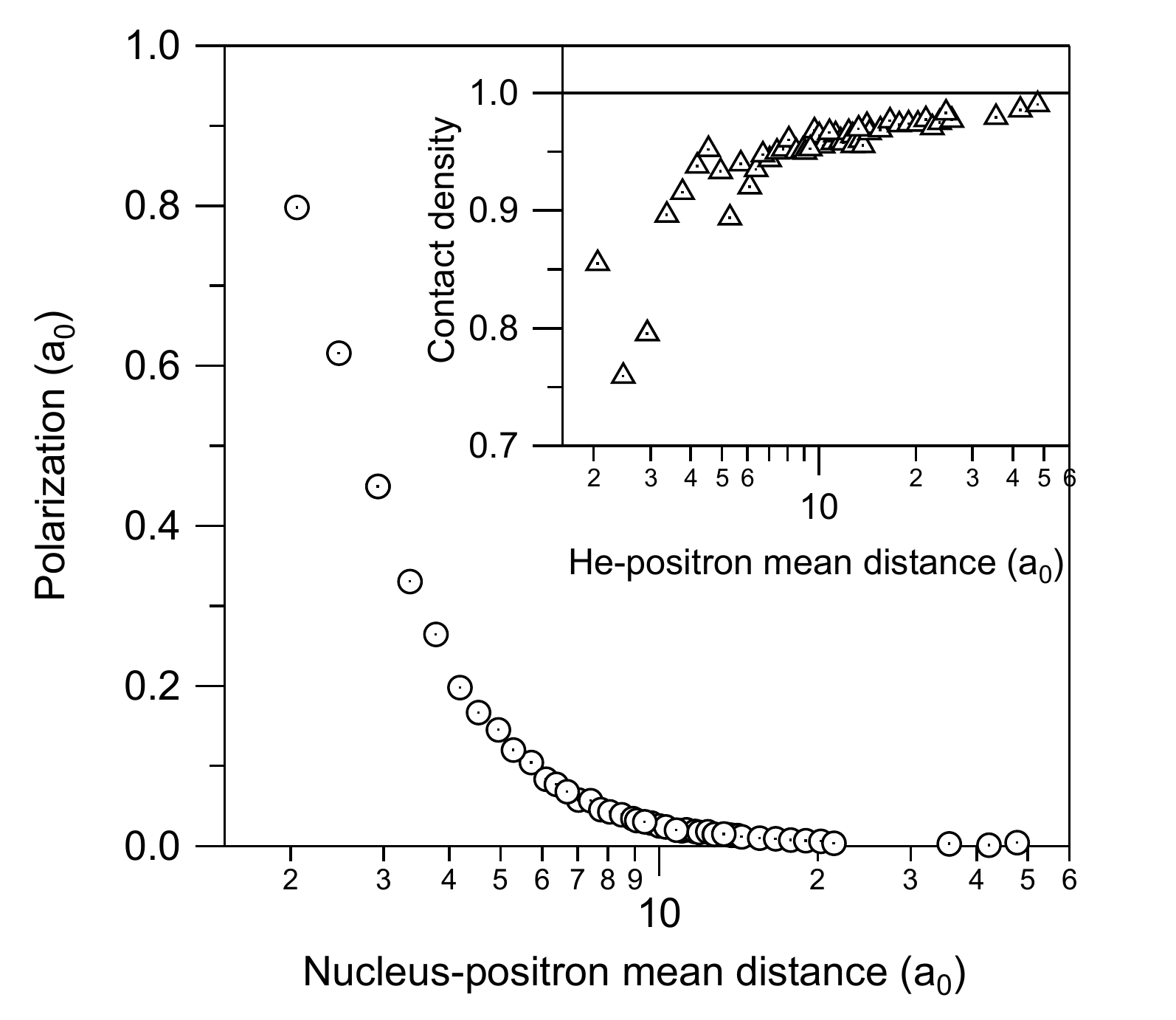}
\caption{The polarization parameter for all the calculated
HePs configurations as a function of the positron mean distance \rp{}
from the He nucleus. The inset shows the electron-positron contact density 
corresponding to the Ps atom component as a 
function of \rp{}. The solid line indicates the contact density of free Ps.}
\label{fig9}
\end{center}
\end{figure}

\begin{figure}[ht]
\begin{center}
\includegraphics[width=8.5cm]{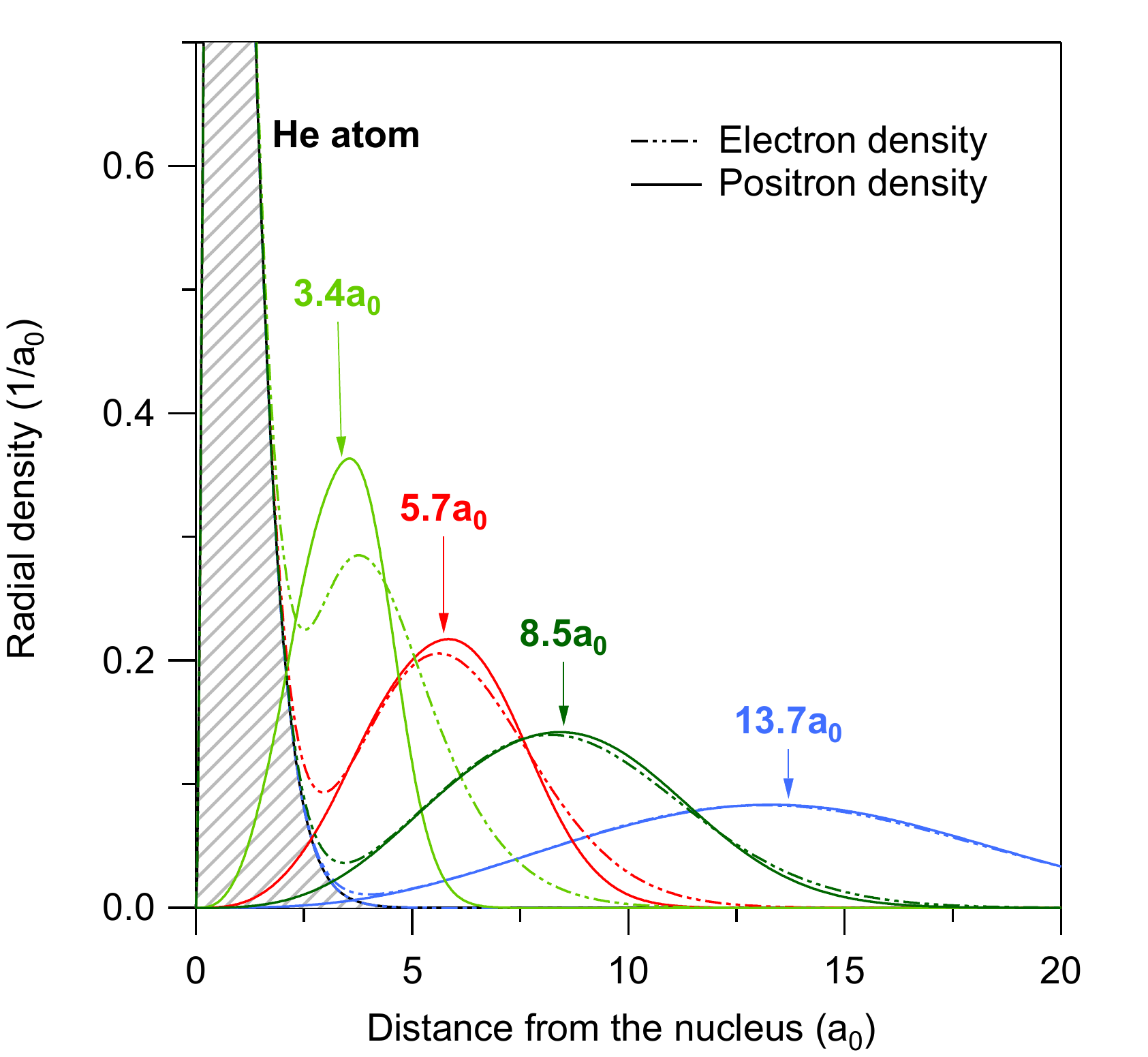}
\caption{(Color online) Radial distributions of the positron (full
lines) and electrons (dashed  lines) for four different configurations 
of the HePs system with \rp{} values of 3.1~\au{} (light green), 5.2~\au{} (red),
7.5~\au{} (dark green) and 12.1~\au{} (blue). 
The arrowed tags indicate the position and the value 
of the positron mean 
distances \rp{} from the He nucleus for each configuration. The shading gives
the electron density of a free He atom. All the distributions are normalized to 
the total number of particles. 
}
\label{fig1}
\end{center}
\end{figure}

Using \rp{} and assuming that the Ps atom is not polarized, \re{} can be 
approximated~\cite{MSF_Zubiaga} 
as the sum of two thirds of the mean electron distance in the
He atom of 0.93~\au{} and one third of the mean positron distance \rp{} 
(blue line in figure~\ref{fig5}). 
The agreement with the \re{} calculated from the 
many-body wavefunction is very good when \rp{} is larger than 7-8~\au{} (or
the nucleus-positron distance parameter $l_{He-p}$ is larger than 4-5~\au{}). 
For smaller $l_{He-p}$ values the actual \re{} is larger than the estimated
one which reflects the polarization of the Ps atom when approaching the He
nucleus. The difference can actually be taken as a measure of the 
polarization (see figure~\ref{fig9}). 
For \rp{}=20~\au{} the polarization parameter
is very small ($\sim$6$\times$10$^{-3}$~\au{}). The polarization starts
to increase when \rp{} is below 10~\au{} and at 4~\au{} its value is $\sim$0.2~\au{}
, i.e., 10\% of the electron-positron mean distance for an isolated Ps. 
The maximum value of the polarization parameter is 0.8~\au{}  
when the interaction energy is 0.44 Ha and Ps is actually unstable against 
dissociation. 
According to our calculations, the smallest stable configuration corresponds to
 \rp{}~$\sim$~2.9~\au{} and a polarization parameter of $\sim$0.45~\au{}. 

The electron-positron contact density of the Ps atom 
(shown in the inset of figure~\ref{fig9}) has been obtained 
by subtracting the contribution of the He electrons
estimated as two times the electron-positron contact density of the $S$=0 
electron-positron pair. 
At \rp{}=20~\au{}, the contact density within the Ps atom 
reaches the value of 0.97, which is close to the exact value of 1 for 
an isolated Ps. At \rp{}=4~\au{} the contact density is still 
$\sim$0.94 and at the smallest separation decreases down to 0.76. The decrease
of the contact density reflects and increasing polarization of the
Ps atom in the interaction with the He atom.

Figure~\ref{fig1} shows the radial distributions of the positron and 
the electrons in HePs configurations corresponding to the positron 
mean distance \rp{} ranging from 3.4~\au{} to 13.7~\au{}. 
The electron distribution can be decomposed into two components. The  
larger corresponds closely to the undisturbed electron density of a He atom.
The second maximum overlaps with the positron distribution especially
well for large \rp{} indicating the formation of a 
well-distinguishable Ps atom.
At small separations, the overlap of the electron and positron distributions
weakens, indicating the increase of the polarization of the Ps atom. 

\begin{figure}[ht]
\begin{center}
\includegraphics[width=8.5cm]{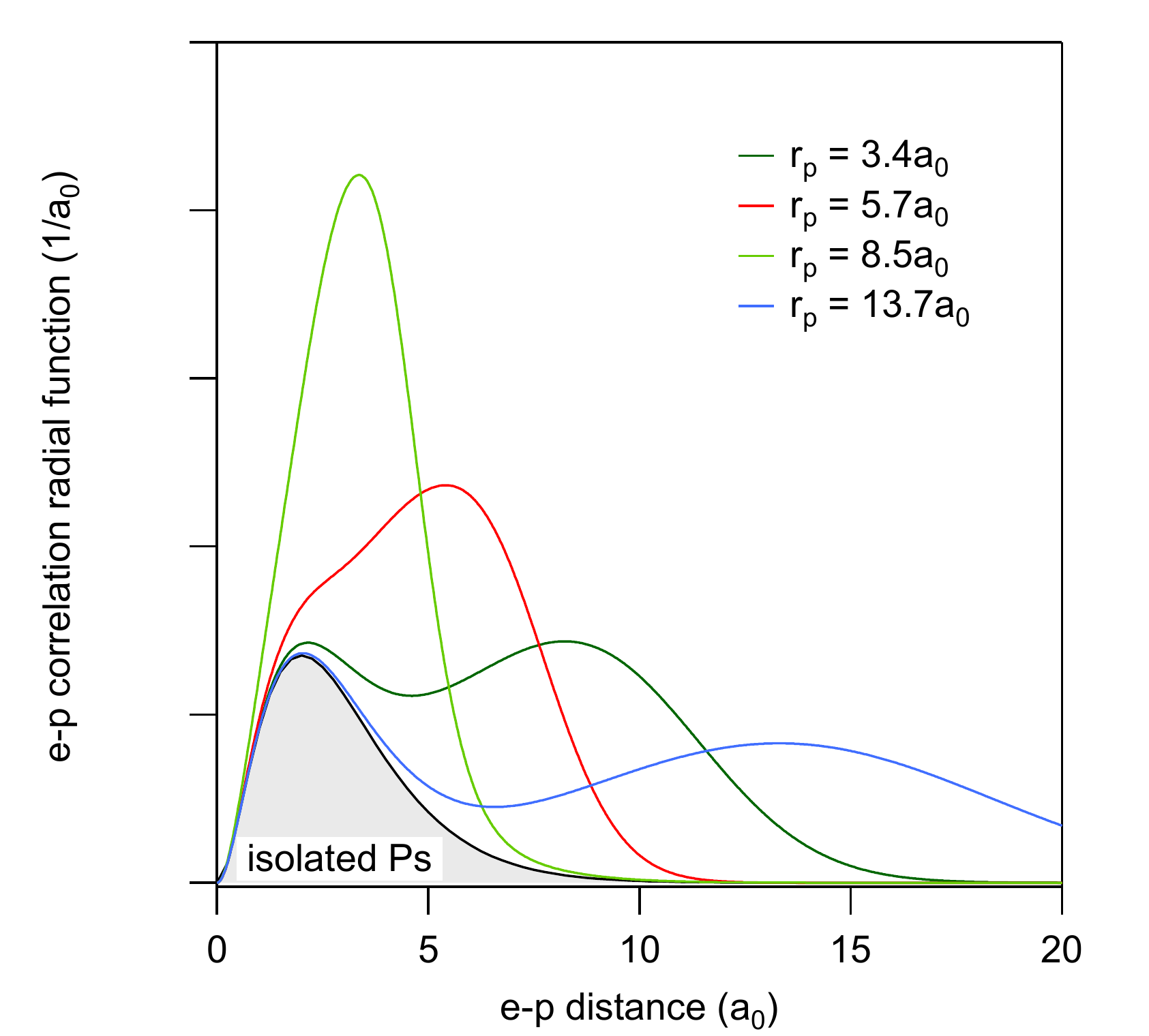}
\caption{Electron-positron pair correlation function for the four HePs 
configurations of figure~\ref{fig1}. The correlation distance for the isolated Ps atom 
(the dashed filled curve) is shown for comparison. All the electron-positron
correlation curves are normalized to the number of electrons in the system
in question. 
}
\label{fig2}
\end{center}
\end{figure}

Figure~\ref{fig2} shows the radial electron-positron correlation distribution
for the four HePs configurations and for the isolated Ps atom. 
The distribution is calculated from equation~(\ref{eq_ep})
and it is normalized to the total number of electrons.
When \rp{} = 3.4~\au{}, the electron-positron pair correlation function shows a single peak 
but when \rp{} is above 5~\au{} a second peak starts to appear, as a shoulder, at 2~\au{}. 
The shape and the magnitude 
of the second peak approach clearly to those for the isolated Ps atom and 
it corresponds to the electron-positron correlation inside the Ps atom. 
The second peak follows the positron distribution when it recedes from the He nucleus 
and it corresponds to the correlation between the positron and
the electrons of the He atom. 
However, even at 8.5~\au{}, the interaction between He and Ps affects 
the electron-positron correlations 
and the appearance of the Ps atom is not so 
clear-cut as in the electron distribution in figure~\ref{fig1}.

\begin{figure}[ht]
\begin{center}
\includegraphics[width=8.5cm]{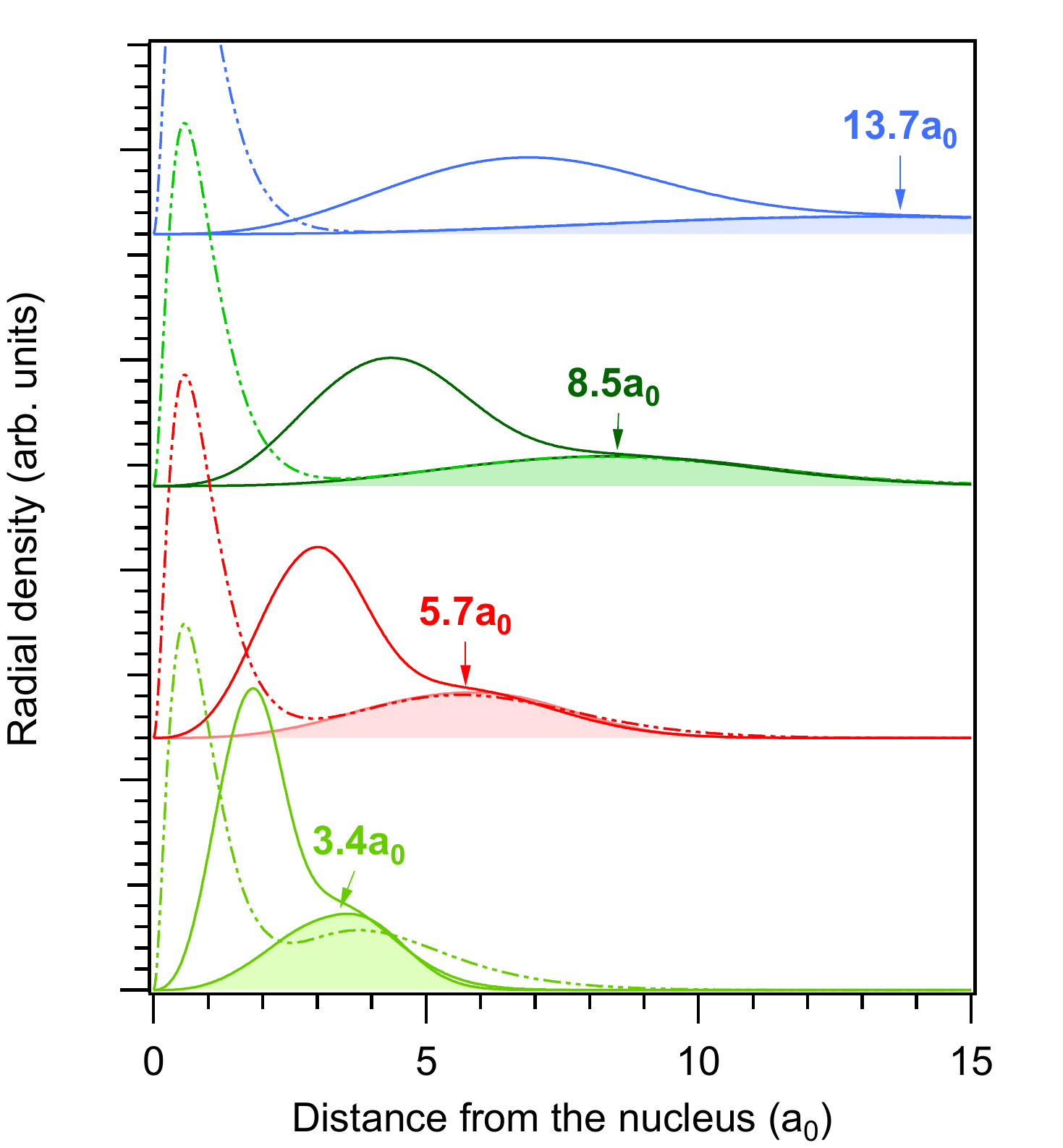}
\caption{(Color online) Electron-positron CM distribution 
for the four HePs configurations. The electron
(dashed) and the positron (filled areas) radial distributions are also
shown for comparison. The CM distributions are normalized to the
number of electrons. }
\label{fig3}
\end{center}
\end{figure}

The distribution of the e-p CM, shown in figure~\ref{fig3}, can also be 
decomposed into two components. The larger component is closer to the
He nucleus and its integrated intensity is double of that for the
smaller and more delocalized component.
Thus, the larger and the smaller components correspond to the 
two electrons of the He atom and the electron of the Ps atom,
respectively. The Ps component is very similar to the 
positron radial density, especially when the mean He nucleus-positron 
distance is larger than 5~\au{}.

\begin{figure}[ht]
\begin{center}
\includegraphics[width=8.5cm]{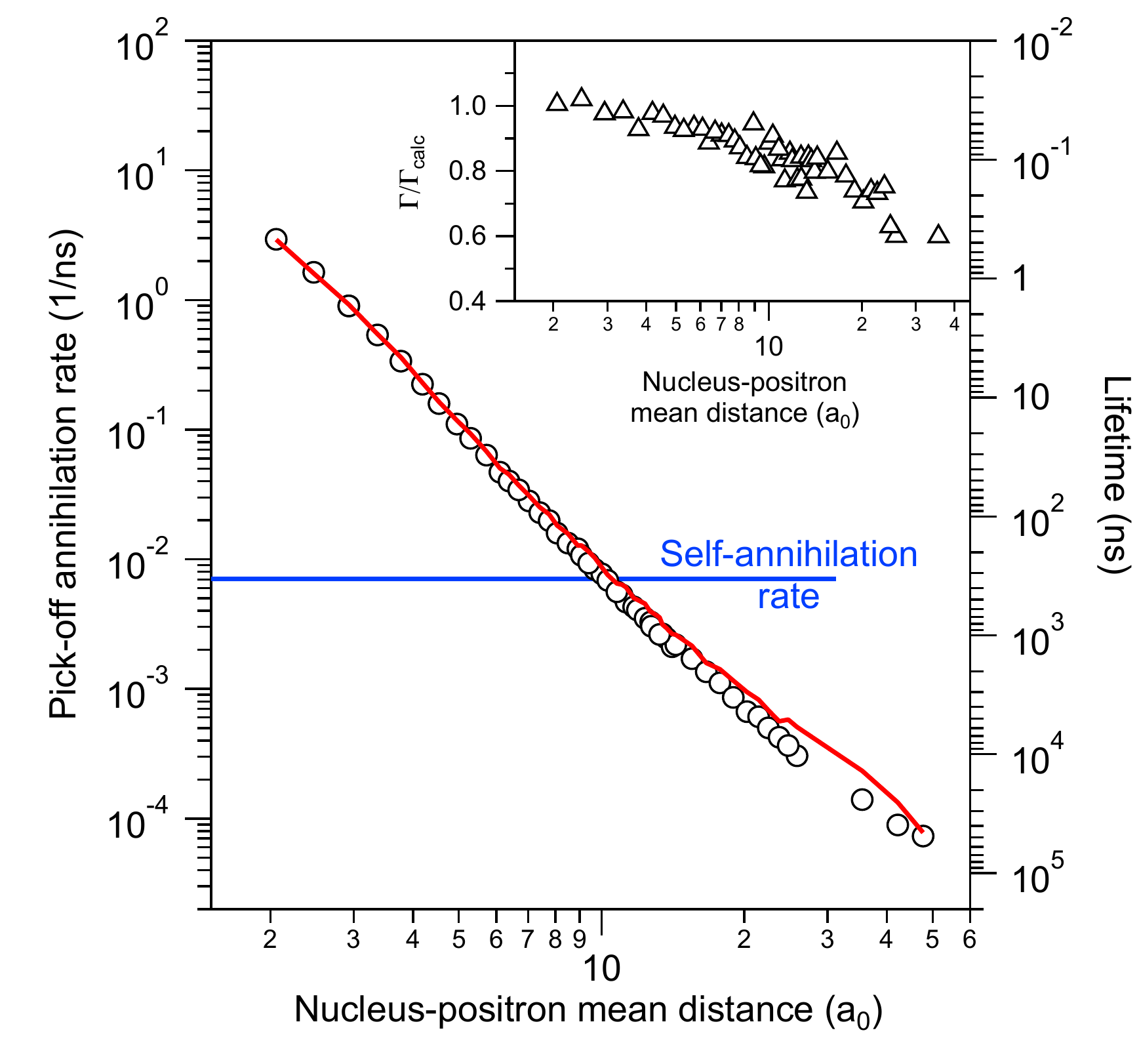}
\caption{(Color online) Positron annihilation rate obtained from the SVM results using 
equation~\ref{gamma_eq} (black circles) and from the overlap of the free He atom and the 
positron densities, using equation~\ref{gamma_calc} (solid line). The inset show the ratio of the values.
The blue horizontal line gives the self-annihilation rate of o-Ps.}
\label{fig11}
\end{center}
\end{figure}

Figure~\ref{fig11} shows the positron pick-off annihilation rate $\Gamma^{po}$ as a 
function of \rp{}, calculated using equation~\ref{gamma_eq}. 
It is an important magnitude for understanding Ps lifetime experiments in molecular matter and 
it also reflects the interaction of the Ps and He atoms and its behaviour. 
The projector $\widehat{P}^{i-p}_{S=0}$ in equation~\ref{gamma_eq} ensures 
that the pick-off processes occur only with the He electron forming a spin-singlet state with the positron. 
$\Gamma^{po}$ increases with the increasing overlap of the 
positron and He electron density.  At \rp{}=10~\au{}, $\Gamma^{po}$
is about the same as the self-annihilation rate 
$\sim$7.04$\times$~10$^{-3}$~1/ns of o-Ps  (lifetime $\tau$ $\sim$ 142~ns). 
At 4~\au{}, $\Gamma^{po}$ $\sim$ 0.2 1/ns ($\tau$ $\sim$ 4.8 ns) and for 
\rp{} $\sim$ 2.5~\au{}, $\Gamma^{po}$ equals to 2 1/ns ($\tau$=500 ps), the spin-averaged annihilation rate of Ps.

A non-selfconsistent positron pick-off annihilation rate is obtained from the
overlap of the free He electron density $n_{e^-}^{He}$ and 
the SVM positron density $n_{e^+}$ as
\begin{equation}
\Gamma_{ov}^{po} = 4\pi r_0^2 c \int d\vec{r}\ \frac{1}{2}n_{e^-}^{He}(\vec{r})
\ n_{e^+}(\vec{r})\label{gamma_calc}.
\end{equation}
The free He electron density is from a SVM calculation and the factor one half reflects 
that only one electron in He annihilates with the positron.
$\Gamma_{ov}^{po}$ is widely used to calculate the pick-off annihilation rate of o-Ps~\cite{PR_Brandt}. 
With increasing nucleus-positron mean distance, it 
agrees with the many-body $\Gamma^{po}$ result  until \rp{}$\sim$5~\au{} 
(solid line in figure~\ref{fig11}). 
When \rp{} further increases, $\Gamma_{ov}^{po}$ remains higher than $\Gamma^{po}$.
The ratio $\Gamma^{po}/\Gamma_{ov}^{po}$, 
shown in the inset of figure~\ref{fig11}, ranges between 1 and 0.6. 
The most confined systems show the largest ratios and it decreases steadily
until \rp{} $\sim$ 30~\au{}. At the level of the o-Ps self-annihilation
rate the overlap of the annihilating positron with the He electrons has 
decreased by 20\% compared to the undistorted He electron density. 
It should be noted that at large separations the overlap and the 
annihilation rate are very small, which is also reflected in the 
scatter of the overlap ratios. However, small values down to o-Ps 
self-annihilation rate are expected to be important for large voids. 

In contrast to a positron in an electron gas, where the positron-electron contact 
density is enhanced~\cite{RMP_Puska}, for Ps interacting with He the many-body effects 
induce a depletion of the contact density. 
The electron in Ps screens 
the charge of the positron and the electron-electron Pauli repulsion 
further decreases the contact density at the positron position, specially 
at low electron density values. Thus, in strongly confined systems the 
contact density is relatively weakly affected, but in weak 
confinement it is strongly depleted. 
$\Gamma_{ov}^{po}$ needs to be corrected for this effect 
 specially when calculating the annihilation rate 
in large voids or pores, where the contact density is small and the annihilation rate is 
largely overestimated. 

\begin{figure}[ht]
\begin{center}
\includegraphics[width=8.5cm]{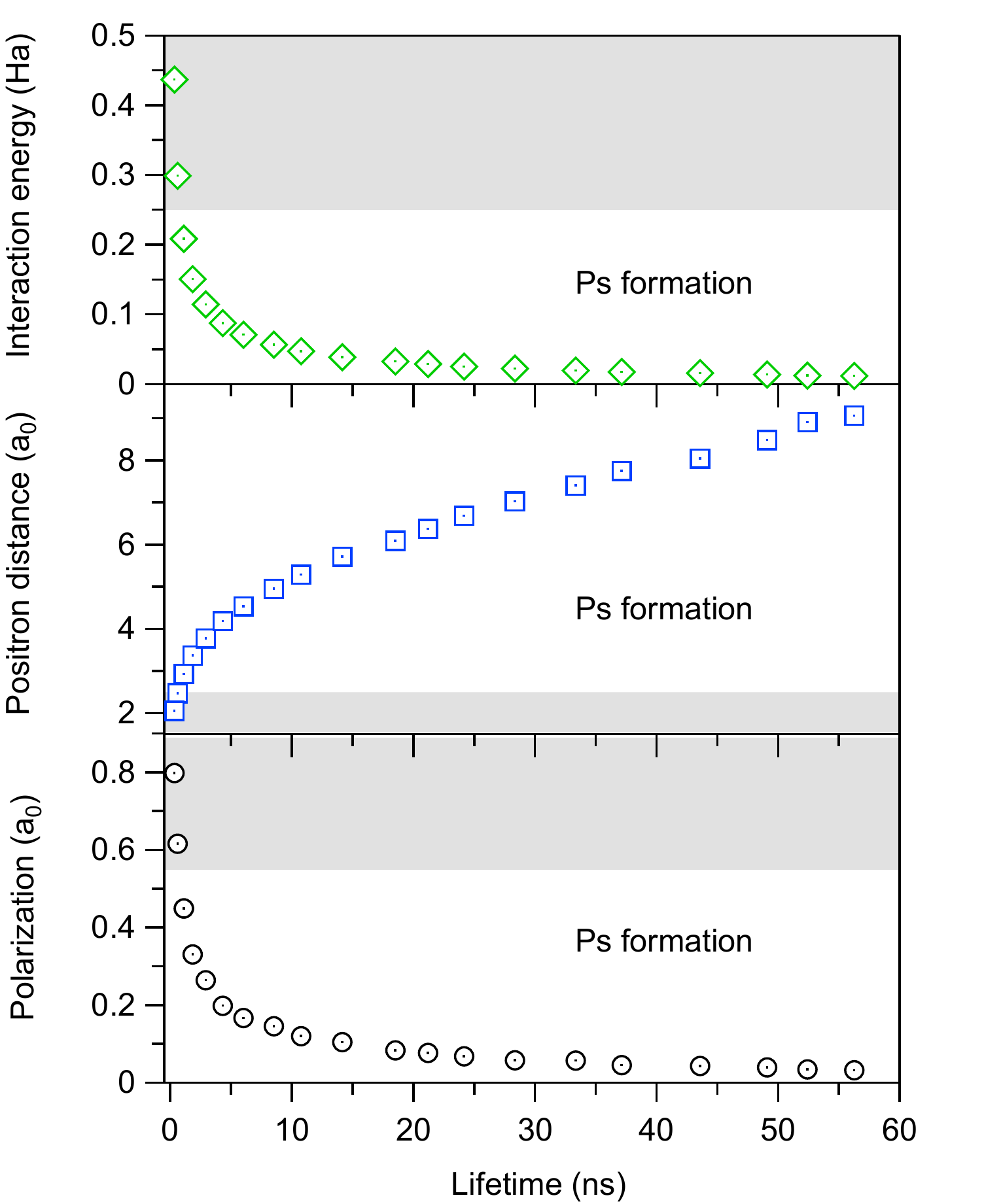}
\caption{(Color online) The polarization (black circles), the He nucleus-positron distance 
(blue squares) and the interaction energy (green diamonds) of the HePs 
system as a function of the positron lifetime. The grey area indicates the value 
range where the interaction energy of Ps atom is larger than its binding energy.
}
\label{fig10}
\end{center}
\end{figure}

Figure~\ref{fig10} resumes the polarization parameter, He nucleus-positron 
mean distance and the interaction energy of the HePs system as a function 
of the positron lifetime (1/$\tau$=$\Gamma^{po}$+1/142.05). The gray area marks 
where the interaction energy of Ps is larger than the Ps binding energy. The figure gives a clear
idea about the magnitude of these quantities in different conditions. 

\section{Conclusions}
We have been interested in to what extent a separate Ps atom can be 
distinguished in a close interaction with a He atom, i.e., when the
volume of the HePs system is constrained. According to our many-body calculations 
the behavior of several measures, i.e., those of electron and positron
mean distances from the He nucleus, electron and positron densities,
as well as electron-positron pair correlation function and center-of-mass 
distribution show the formation of a Ps atom that is weakly polarized until the mean 
He nucleus-positron mean distance \rp{} is shorter than $\sim$5~\au{}. 
At shorter distances the Ps atom also forms but it is strongly polarized.
The most clear-cut evidence is given, simply, by the electron and positron 
densities shown figure~\ref{fig1}. There, the  electron density can be decomposed
into the He and Ps atom components and the positron density overlaps well
with the electron density. 
Even at \rp{} $\approx$ 5~\au{}  the positron-electron contact density 
is still high, $\sim$93~\% of that in the free Ps atom,
and the polarization parameter indicating the imbalance of the
electron and positron densities in the Ps atom is small, 0.2~\au{}.

The above notions for the HePs system, generalized to molecules, supports
the possibility to use a practical single-particle description for o-Ps 
in molecular materials. Many-body calculations could be used to define 
pairwise atom-Ps potentials which could form the basis for the calculation 
of the Ps energy landscape in the material and obtain, subsequently, its 
distribution and pick-off annihilation lifetime. In that respect, the 
above-mentioned findings about the relative depletion of the contact density, 
will simplify the reliable estimation of the pick-off annihilation rate.

\begin{acknowledgments}
This work was supported by the Academy of Finland through the postdoctoral researcher fellowship and the centre of excellence programs. Thanks are due to K. Varga for proving us his the ECG-SVM code. We also thank I. Makkonen, J. Mitroy and T. Rantala for helpful discussions. The comments and suggestions from I. Makkonen have been of great help while preparing the manuscript. 
\end{acknowledgments}

\end{document}